\documentclass[preprint]{imsart}
\RequirePackage[OT1]{fontenc}
\RequirePackage{amsthm,amsmath}
\RequirePackage[numbers]{natbib}
\RequirePackage[colorlinks,citecolor=blue,urlcolor=blue]{hyperref}
\usepackage{graphicx}
\usepackage[figuresright]{rotating}
\usepackage{amsmath}
\usepackage{times}
\usepackage{amsfonts,amsbsy, amssymb,float, mathtools,multirow, url}
\usepackage{bm}
\newcommand{\bs}[1] {\bm{#1}}
\usepackage{natbib}
\usepackage{geometry}
\usepackage{url}
\usepackage{hyperref}       
\usepackage{cleveref}       
\crefname{algocf}{alg.}{algs.}
\Crefname{algocf}{Algorithm}{Algorithms}
\usepackage{url}            
 
\usepackage{float}
\usepackage{rotating}
\usepackage[ruled,linesnumbered]{algorithm2e}
\usepackage{xcolor}
\usepackage{tikz}
\usepackage{booktabs}
\usepackage{comment}

\startlocaldefs
\newcommand{\wang}{\textcolor{red}}

\numberwithin{equation}{section}
\theoremstyle{plain}
\newtheorem{theorem}{Theorem}[section]
\newtheorem{Proposition}{Proposition}[section]

\newtheorem{Assumption}{Assumption}

\DeclareMathOperator*{\argmax}{arg\,max}
\DeclareMathOperator*{\argmin}{arg\,min}

\DeclareMathOperator*{\E}{\rm E}
\DeclareMathOperator*{\Pro}{\mathbb{P}}

\endlocaldefs 
\begin{document}

\begin{frontmatter}

\title{Empowering Multi-class Classification for Multivariate Functional Data with Simultaneous Feature Selection}
\runtitle{Complex Functional Data Classification with Feature Selection via Deep Learning}
\begin{aug}
\author{\fnms{Shuoyang} \snm{Wang} \ead[label=e1]{shuoyang.wang@louisville.edu}}
\and
\author{\fnms{Guanqun} \snm{Cao}\thanksref{t2} \ead[label=t2]{caoguanq@msu.edu}}
\and
\author{\fnms{Yuan} \snm{Huang} \ead[label=e2]{yuan.huang@yale.edu}}
\address{
Department of Bioinformatics and Biostatistics, University of Louisville, U.S.A. \\
\printead{e1}
}
\address{
Department of Statistics and Probability, Michigan State University, U.S.A. \\
\printead{t2}
}
\address{ 
Department of Biostatistics, Yale University, U.S.A. \\
\printead{e2}}
 
\thankstext{t2}{To whom correspondence should be addressed.}
\runauthor{Wang et al.}

\end{aug}

\begin{abstract}
{The opportunity to utilize multivariate functional data types for conducting classification tasks is emerging with the growing availability of imaging data. 
Inspired by the extensive data provided by the Alzheimer's Disease Neuroimaging Initiative (ADNI), we introduce a novel classifier tailored for multivariate  functional data. Each observation in this framework may be associated with numerous functional processes, varying in dimensions, such as curves and images.
 Each predictor is a random element in an infinite dimensional function space, and the number of functional predictors $p$ can potentially be much greater than the sample size $n$. In this paper, we introduce a novel and scalable classifier termed functional BIC deep neural network. By adopting a sparse deep  Rectified Linear Unit   network architecture and incorporating the LassoNet algorithm,  
   the proposed unified model  performs feature selection and classification  simultaneously, which is contrast to the existing functional data classifiers.  The  challenge arises from the complex inter-correlation structures among multiple functional processes, and at meanwhile without any assumptions on the distribution of these processes.
  Simulation study and real data application are carried out to demonstrate its favorable performance.}
\end{abstract}

\end{frontmatter}

\section{Introduction}
\label{s:intro}
  Data considered in functional data analysis (FDA) include curves and images, which are commonly viewed as infinite-dimensional random vectors in a  function space.
Compared with classical 
 independent and identically distributed (i.i.d.) data, the key distinguishing feature of functional data is the presence of dependence and smoothness within each data curve.   
There has been a recent surge in applications of multivariate functional data analysis due to new developments in neuroimaging (e.g. fMRI and PET)  and electroencephalogram. 
Conventionally, the  functional data is a collection of independently and randomly observed samples which are real-valued functions, $\left\{ X_i(\bm s)\right\}_{i=1}^n$, where $\bm s\in\mathcal{S}$, is defined on a compact interval $\mathcal{S}\subset \mathbb{R}^d$, $d\geq 1$. When $d=1$, it refers to classical 1D functional data, i.e., curve data. When  $d> 1$, it refers to multi-dimensional functional data, such as 2D or 3D imaging data ($d=2,3$).

Our motivating example stems from the Alzheimer's Disease (AD) Neuroimaging dataset (\url{http://adni.loni.usc.edu}), which is a longitudinal multicenter study designed to develop clinical, imaging, genetic, and biochemical biomarkers for early detection and tracking of AD. Given the dataset's multidimensional nature, particularly its incorporation of PET imaging data and biochemical biomarker curve data, there exists significant scientific interest in predicting the status of new patients, whether they belong to the AD group, early mild cognitive impairment  (EMCI) group, or the control (CN) group.
The data's composition, featuring multiple imaging and curve data, lends itself naturally to FDA, which has emerged as a popular method for AD data analysis and classification \citep{Wang:etal:14:AOAS,Wang:etal:16:bios,Li:etal:22}. Notably, this dataset's functional representation allows for a deeper understanding of the disease's progression and offers valuable insights into potential diagnostic markers.
Furthermore, considering the potential correlations among the observed functional data, we posit that a novel functional classifier, capable of capturing these correlations while selecting significant functional features, would be immensely advantageous in this multivariate functional data setting. 
{As such, an approach that can capture  correlations among signals and identify key functional features essential for accurate classification is needed.} 

\subsection{Related works}

There has been rich literature on functional data classification with a scalar response and univariate functional predictors. The prevalent technique applied in functional classification is based on the density function, the regression model and distance function.  \cite{Muller:05classification} considered the generalized functional linear model for the sparse functional data, where the integral is estimated by the first several truncated projection scores. \cite{James:Hastie:01}   considered the representation of functional observations using splines basis, and they applied multivariate linear discriminant analysis  to the spline coefficients to construct the classifier.    \cite{Delaigle:Hall:12} proposed the functional centroid classifier, where the distance is constructed by the absolute difference of inner products. \cite{Cerou:06}  explored the nearest neighbor classifier in infinite dimensional metric space, and argued that the weak convergence of nearest neighbor classifier holds only if the metric space is separable with some regular conditions. Current functional data classification approaches often assume that the functions are i.i.d., limiting their ability to classify subjects based on multiple observed functions that are expected to be correlated. Recently, efforts haven been made for multivariate functional data classification.   \cite{Dai:Genton:18} proposed classifiers  for both univariate and multivariate functional data by extending the depth-based scalar outlyingness to an outlyingness matrix. \cite{moindjie:etal:22} considered partial least square classification   method for multivariate functional data  defined in different domains.

As a more powerful tool to characterize the  complex functions,   neural network  methods are recently used for   FDA. For instance, \cite{yao2021deep} introduced   adaptively learned bases that are implemented via micro neural networks. \cite{rao2023nonlinear} proposed a continuous layer that preserves
the functional characteristics of the data. \cite{thind2023deep,wu2023neural} introduced a functional autoencoder architecture tailored for    functional input and output.
Deep neural network (DNN) methods are recently used for multi-dimensional  functional data classification to represent the powerful and complex classifiers. In particular, \cite{wang:etal:21a}   and \cite{wang:etal:23} proposed an optimal DNNs classifier for both  densely and sparsely observed multi-dimensional functional data.
However,  they remain  challenging to understand the dependency among multivariate  functional data. This is either because their functional representations lack the flexibility to capture intricate features or because they are designed  for either i.i.d. curve functional data only \cite{wang:etal:21a}, or i.i.d. imaging data only \cite{wang:etal:23}.  
For a more detailed discussion of the development of functional data classification, we refer to recent review papers such as  \cite{Wang:Huang:cao:23}.



Despite the emphasized importance of  feature selection in the FDA literature, limited research has been conducted on analyzing their impacts on functional data classification. In existing literature, the focus of feature selection applied to functional data primarily revolves around replacing each infinite-dimensional observation with a finite-dimensional vector.
For instance, \cite{Berrendero:etal:16} employed variable selection to identify intervals of the functional inputs corresponding to local maxima of the distance-covariance function. Similarly, \cite{Yu:etal:22} proposed Gaussian process discriminant analysis, integrating variable selection and classification by treating each  1D functional data as a high-dimensional input and selecting significant intervals.

\subsection{Our contributions}

To address the aforementioned challenges and accomplish the classification task outlined in the motivating example,  we introduce a novel and scalable classifier termed functional BIC deep neural network (fB-DNN). This classifier seamlessly integrates feature selection and classification for multivariate multi-dimensional functional data.

We adopt a sparse deep  Rectified Linear Unit (ReLU) network architecture and incorporate the LassoNet algorithm proposed by \cite{Lemhadri2018}. LassoNet, a neural network architecture combining Lasso regularization with deep learning techniques, is designed to  perform feature selection for supervised learning. Despite its demonstrated effectiveness in sparse DNNs for cross-sectional high-dimensional data settings, its potential in the context of functional data classification remains under-explored.
We adapt the LassoNet algorithm from \cite{Lemhadri2018} to suit the requirements of functional classification, while also incorporating a high-dimensional BIC criterion to address the tendency of the original LassoNet algorithm to retain redundant features based on our empirical observations. Additionally, we develop the universal approximation theory and establish global optimality for the proposed classifier under the multivaraite multi-dimensional functional data setting. We have also developed the convergence rate of the proposed classifier to the Bayes error when the optimizer has been reached.

Compared with the existing studies, we have two significant contributions. First, the DNN architecture we propose is fundamentally different. In this work, we introduce a novel DNN class that incorporates residual layers and penalty functions specifically for functional data. This architecture allows for functional feature selection. 
The inclusion of these components not only enhances the model's ability to perform classification and feature selection simultaneously but also presents additional computational challenges.
Second,  including feature selection in functional data classification  enhances interpretability and reduces monitoring costs by considering fewer functional varieties capable of discriminating the behavior of functional data. Feature selection leads to improved classifier performance by mitigating the negative impact of correlations between features on classification rates.
For instance, in ADNI data analysis, classification accuracy is compromised when all functional features (such as 2D imaging slices) are included, as detailed in Section \ref{SEC:realdata}. Thus, selective inclusion of features based on their  discriminative power can be a valuable feature. 

This article is organized as follows. In Section  \ref{SEC:formulation}, we demonstrate the proposed functional classifier and feature selection procedure. An implementation of the the  proposed classifier is considered in Section \ref{sec:implement}. Simulation studies   are conducted in Section  \ref{sec:sim}. Section \ref{SEC:realdata} gives the application to the motivating real data. The conclusion of our study is summarized in Section \ref{SEC:summary}. All proofs and additional numerical analysis results are given in the Appendix.

\section{Formulation of  functional data classification}\label{SEC:formulation}

\subsection{Model setup}

Throughout the paper, we consider $p$-dimensional functional covariates $\bm X = \left\{X_j\left(\bm s_j\right) \right\}_{j=1}^p$. In general, $X_j$'s are random processes defined on various domains with different dimensions $d_j$. Without loss of generality,  we assume $\bm s_j\in\left[0,a_1\right]\times \cdots\times\left[0,a_{d_j}\right]$ and $a_l>0$, $l=1,\ldots,d_j$. Let $K\geq 2$ be the number of classes, and the class label $Y\in\left\{1,\ldots, K \right\}$. For the class $k$, we denote the prior $\pi_k = \Pro\left(Y=k \right)$, such that $\sum_{k=1}^K \pi_k = 1$. In addition, we define the inner product $\langle\cdot,\cdot\rangle$, such that for any domain ${\mathcal{S}}$, $\langle f, g\rangle  = \int_{\mathcal{S}}f(\bm s)g(\bm s)d\bm s$, where $f,g\in L^2\left( \mathcal{S}\right)$.

Suppose we observe $n$ i.i.d. functional training samples $\left\{(\bm X_i, Y_i): 1\le i\le n\right\}$, which are independent of $\bm X$ to be classified. We are interested in grouping a newly observed set of functional covariates $\bm X$  to one of the $K$ classes given the training data. Notably, each $X_j: \left[0,1\right]^{d_j} \rightarrow \mathbb{R}$ is assumed to be in $L^2\left(\left[0,1\right]^{d_j}\right)$, such that $\E\langle X_j(\bm s_j), X_j(\bm s_j) \rangle<\infty$.  {Given $Y=k$, $X_j\left(\bm s_j\right)$ presumably 
 has some unknown mean function $\mathbb{E}X_j\left(\bm s_j\right)= \mu_{j|k}\left(\bm s_j\right)$ and unknown covariance function $$\Omega_{j|k}(\bm s_j,\bm s'_j)=\mathbb{E}\left[\left(X_j(\bm s_j)-\mu_{j|k}(\bm s_j)\right)\left(X_j(\bm s'_j)-\mu_{j|k}(\bm s'_j)\right)\right].$$ 

In the following, we assume that there exists a class of nonlinear discriminant functions $\left\{f_k\right\}_{k=1}^{K}$ of $\bm X$, such that the class label $Y$ is assigned to class $k'$ if $k' = \argmax_{k=1,\ldots,K} $ $f_{k}\left(\bm X\right)$.  
This  non-parametric approach accommodates the intricacies of functional data, employing DNNs to navigate the classification conundrum. The novelty of this strategy lies in its resilience against the unpredictable nature of functional data distributions, representing a substantial advancement over traditional methods constrained by rigid model specifications. 

\subsection{Deep neural network classifier}\label{sec:DNN classifier}
{We denote the proposed function class $\mathcal{D}$, whose elements are functional classifiers taking $\bm X$ as input and $\left\{1,\ldots, K\right\}$ as output. 
We construct the nonlinear transformation $\{H_{j,\ell}\}$ in the first layer by 
$$H_{j,\ell}(X_j) = \sum_{t=1}^{q_0-1} W_{0, t\ell}\sigma\left(W_{0, q_0\ell} + \int_{\left[0,1\right]^{d_j}}X_j(\bm s_j)\phi_{j, \ell}(\bm s_j)d\bm s_j\right),  $$
where $W_{0, t\ell}$ is the $(t,\ell)$th entry of weight matrix $W_0\in\mathbb{R}^{q_0\times q_1}$, $q_0$ is the number of nodes in the initial layer transforming the functional inputs, $q_1$ is the number of nodes passed to the next layer, and the last row $W_{0, q_0\ell}$ is the shift vector. Note that the proposed first layer encode the infinite-dimensional functions $X_j(\bm s_j)$ to some finite number of numerical neurons by introducing functional weights $\left\{\phi_{j, \ell}\right\}_{\ell=1}^{q_{1j}}$ for bridging the input and the first hidden layer, where $q_1=\sum_{j=1}^p q_{1j}$.  Each weight function   $\phi_{j, \ell}: \left[0,1\right]^{d_j} \rightarrow \mathbb{R}$ is assumed to be within $L^2\left(\left[0,1\right]^{d_j}\right)$. Let $\sigma$ be the activation function, such as ReLU. 
Practically, the selection of $\phi_{j, \ell}$ and $q_1$ 
plays a crucial role in the classification performance, and we will discuss the details in Section \ref{sec:implement}.  
}

Next, we continue to construct the class of residual feed-forward neural networks. These networks are not only noted for the ease of training \citep{He2016}, 
but are also acclaimed for their versatility in serving as universal approximators for a diverse array of function classes \citep{Lin2018}. Let $\bm W$ be a collection of weight matrices $W_l\in\mathbb{R}^{q_l\times q_{l+1}}$, $l=1,\ldots, L$. Similar to the definition of $H_{j, \ell}$, for the $(l+1)$th hidden layer, there are $q_{l+1}$ nodes and  $q_l$ nodes in the previous layer. The residual feed-forward neural networks is defined as
 \begin{equation}\label{dnn class}
    \mathcal{D} = \left\{f_{\bm b, \bm W} : f = \sigma^\ast\left(\sum_{j=1}^p \bm b_j^\intercal{\bm H}_j^{(q_{1j})} + g_{\bm W}\left({\bm H}_1^{(q_{11})}, \ldots, {\bm H}_p^{(q_{1p})} \right)\right)\right\},
\end{equation}
where ${\bm H}_j^{(q_{1j})} = \left(H_{j,\ell}(X_j), \ldots, H_{j, q_{1j}}(X_j) \right)^\intercal$, $\bm b = \left(\bm b_1^\intercal, \ldots, \bm b_p^\intercal \right)^\intercal$ denotes the weights for residual layer and  $\sigma^\ast$ is an activation function to be selected according to the number of class labels. Any generic  nonlinear function $g$ has the composition representation \citep{Schmidt:19,wang:cao:23}
 {\begin{equation}\label{EQ:g}
g(\bm x) =  {W}_L\sigma {W}_{L-1}\sigma\ldots {W}_2\sigma {W}_1\bm x, \,\,\,\, {\bm x}\in\mathbb{R}^{q_1}.
\end{equation}} 
We estimate the classifier $f$ by  optimizing the following problem: 
\begin{equation}\label{eq:objective}
\widehat{f}=   \argmin_{f\in\mathcal{D}} n^{-1}\sum_{i=1}^n \mathcal{L}\left(Y_i, f\left(\bm X\right)\right),
\end{equation}
where  the loss function $\mathcal{L}$ is chosen regarding the classification task.  

For the binary classification with $K=2$,  one common option for $\mathcal{L}$ is the hinge loss. 
 Consequently, adjusted by the label $Y_i=\left\{1, 2 \right\}$, the loss function $\mathcal{L}$ is defined as
 $\mathcal{L}\left(Y_i, f\left(\cdot\right) \right) = \max\left(1-\left(2Y_i-3\right)f\left(\cdot \right) , 0\right)$.
The optimizer of Equation (\ref{eq:objective}) is essentially the estimate of log-likelihood ratio of the two classes $f^\ast$, thus the activation function $\sigma^\ast$ is the identity function, and the new label will be estimated by the sign of $\widehat{f}$, i.e,
\begin{equation}\label{DEF:Yhat}
{\widehat{Y}}=
\left\{\begin{array}{cc}
2, & \widehat{f}\left(\bm X\right)\ge 0,\\
1, & \widehat{f}\left(\bm X\right)< 0.
\end{array}\right.
\end{equation}

In general, for the multi-class classification with $K\geq 2$,  the activation function  $\sigma^\ast$  is selected as the $K$-dimensional softmax  function $\sigma^\ast(\bm z) = \left(\frac{\exp{(z_1)}}{\sum_{k=1}^K\exp{(z_k)}}\right.$, $\left.  \ldots, \frac{\exp{(z_K)}}{\sum_{k=1}^K\exp{(z_k)}}\right)$ for $\bm z=(z_1, \ldots, z_K)\in \mathbb{R}^K$.   Define the label as $\bm y=\left(y_1,\ldots, y_K \right)^\intercal$, such that $y_k=1$ if $Y=k$ and $0$ otherwise.  The loss function $\mathcal{L}$ is the cross-entropy function
\begin{equation}\label{DEF:crossEn}\mathcal{L}\left(Y_i, f\left(\bm X \right) \right) = -\bm y_i \log\left( f \left(\bm X \right)\right),
\end{equation} 
where $ f \left(\bm X \right) = \left( f_1 \left(\bm X \right), \ldots,  f_K\left(\bm X \right) \right)^\intercal$. Cross-entropy loss is a frequently employed loss function. This metric   measure the difference between the projected probability distribution and the actual probability distribution of the target classes. 
  The cross-entropy loss will be substantial, for instance, if the model forecasts a low probability for the right class but a high probability for the incorrect class. The optimizer $ \widehat{f}_k$ is the estimation of conditional probability density $f_k$ under the $k$-th class, which is between $0$ and $1$ confined by the softmax function. Therefore, the new label is assigned by 
$\widehat{Y} = \arg\max_{\left\{k=1,\ldots,K \right\}} \left\{k: \widehat{f}_k\left(\bm X \right) \right\}$.


The following theorem presents the uniform approximation theory of our proposed neural network class to the Bayesian classifier. For the sake of space limit,  the   function classes $\mathcal{G}$ and $\left\{\mathcal{G}_k\right\}_{k=1}^K$  and related notations $\alpha_1$ and $\alpha_2$ are introduced in the Appendix. We denote $s$ as the number of non-zero weights for the sparse neural network class $\mathcal{D}$.

\begin{Assumption}\label{assumption1}
Given any $\epsilon>0$, the network class $\mathcal{D}$ satisfies
$$L\asymp \log_2(\epsilon^{-1}),\,\,\| \bm q\|_\infty \asymp \epsilon^{-\alpha_1},\,\,
s\asymp \log_2(1/\epsilon) \left(\epsilon^{-\alpha_1}\right) , \|\bm W \|_\infty\asymp \epsilon^{-1},$$
where $\alpha_1$ is a constant depending on $\mathcal{G}$.
\end{Assumption}

\begin{Assumption}\label{assumption2}
Given any $\epsilon>0$, the network class $\mathcal{D}$ satisfies 
$L\asymp \log_2(\epsilon^{-1})$, $\| \bm q\|_\infty \asymp \epsilon^{-\alpha_2},\,\,
s\asymp \log_2(1/\epsilon) \left(\epsilon^{-\alpha_2}\right)$,
where ${\alpha_2}$ is a constant depending on $\left\{\mathcal{G}_k\right\}_{k=1}^K$.
\end{Assumption}

 To better illustrate the following theorem for binary classification ($K=2$),   the boundary set is $\mathcal{B}(\epsilon) = \left\{\bm x: |\Pro\left(Y=1|\bm X=\bm x \right)-1/2|>2\epsilon\right\}$,
where $\epsilon>0$, and $\bm X$  { is encoded by $q_1$ neurons in the first hidden layer.}
Note that this set controls the distance between the regression function and $1/2$, and its size is decreasing with $\epsilon$. When $\epsilon\rightarrow 0$, $\mathcal{B}(\epsilon)$ expands to cover the entire space of functional covariates.

\begin{theorem}\label{THM:thm1}
Given any universal $\epsilon>0$, when the activation function $\sigma$ in (\ref{EQ:g}) is ReLU, it follows that: 
\noindent{\it(a) } For $K=2$ and identity function $\sigma^\ast$ in (\ref{dnn class}), if $f_1,f_2\in\mathcal{G}$, 
    there exists $\widehat{f}\in\mathcal{D}$ satisfying Assumption \ref{assumption1}, such that 
    $\widehat{f}(\bm X) = f^\ast(\bm X), \bm X\in \mathcal{B}(\epsilon)$,
    where $f^\ast$ is the Bayesian decision function; 
\noindent{\it(b) } For $K\geq 2$ and softmax function $\sigma^\ast$ in (\ref{dnn class}), if $f_k\in\mathcal{G}_k$ and $\min_{k=1,\ldots,K}\|f_k\|_{\infty}\geq \epsilon>0$ , there exist $\widehat{f}_k\in\mathcal{D}$ satisfying Assumption \ref{assumption2}, such that 
    $\|\widehat{f}_k - f_k \|_\infty\leq \epsilon$, $k=1,\ldots,K$.   

\end{theorem}

Theorem \ref{THM:thm1} demonstrates the universal approximation theory using ReLU activation function for both binary and multi-class classification scenarios. Note that $f^\ast$ is essentially the log ratio of probabilities of two classes, and Theorem \ref{THM:thm1} (a) indicates that in any subset of $\mathbb{R}^{q_1}$, with a well selected DNN structure, we are able to construct the  Bayesian decision rule exactly. For multi-class classification with the softmax function, it is crucial to accurately approximate the conditional probability $f_k$, and Theorem \ref{THM:thm1} (b) shows that any $f_k$ can be well approximated with a finely tuned architecture. The proof is provided in the Appendix.

\subsection{Functional feature selection}
Effective variable selection can identify functional covariates that provide significant predictive power.
In the following, we consider a regularization approach with constraints on weights.  This architecture  involves multiple layers of neurons with Lasso regularization applied to the weights connecting the neurons in each layer. During training, the Lasso penalty term is added to the loss function, promoting a sparse solution by shrinking some weights towards zero. This encourages the model to focus on the most informative features while discarding irrelevant ones. Specifically, we estimate  $f(\cdot)$ by minimizing 
\begin{align}\label{eq:obj}
    \argmin_{f\in\mathcal{D}} n^{-1}\sum_{i=1}^n \mathcal{L}\left(Y_i, f\left(\bm X \right)\right)  + \lambda \sum_{j=1}^p P_j\left(\bm b_j\right), 
     \text{s.t. } \|W_1^{(j)}\|_\infty  \leq C \|\bm b_j \|, 
\end{align}
where $P_j(\cdot):\mathbb{R}^{q_{1j}} \rightarrow \mathbb{R}$, $j=1,\ldots, p$, is the penalty function for regularization; the hierarchy coefficient $C$ controls the relative strength of the linear and nonlinear components; the {$L_1$} penalty coefficient $\lambda$ controls the complexity of the fitted model; and $W_1^{(j)}$ is the sub-matrix of $W_1$ connected to $\bm H_{j}^{(q_{1j})}$. The regularizers $P_j(\cdot)$ impose sparsity at the group level for entire $\bm b_j$, meaning that the entire groups of nodes can be shrunk to zero. The estimated $\left\{\bm b_j\right\}_{j=1}^p$ automatically estimates the true sparse set $\mathcal{A} = \left\{j: \| \bm b_j\|\neq 0, j\in\{1,\ldots,p\} \right\}$.
The constraints are designed to calibrate the level of non-linearity attributed to feature $j$. 
If the entire vector $\bm b_j$ are forced to be zero, then all weights connected to the neural network that are related to  $X_j$ are effectively shrunk to be zero, leading to the exclusion of the $j$th feature from the model. This idea extends the LassoNet algorithms proposed in \cite{Lemhadri2018} for i.i.d. high-dimensional data, where $P_j$ are set to be the group-LASSO penalty. 

The selection of the tuning parameters is important in practice.
{Cross-validation is a popular approach and is integrated into LassoNet, making it an intuitive first choice when extending the LassoNet algorithm. We name the associated algorithm  “f-DNN”. See f-DNN algorithm in Section \ref{sec:sim} for details. However, we have observed that its performance for selecting functional features is not satisfactory. }
\cite{Wang:13} recently proposed high-dimensional BIC for  ultra-high dimensional regression when $p$ is much larger than $n$. Let $\bs{\eta}=\left(L, \bm{q}, \gamma, \lambda\right)$ be the tuning parameter set for loss function $\mathcal{L}$, where  $\gamma$  is the dropout rate of neurons. Note that any fixed $(L, \bm{q}, u)$ determines the architecture of the neural network class $\mathcal{D}$. Motivated by \cite{Wang:13}, we
choose tuning parameters $\bs\eta$ that minimizes the following   functional data BIC criterion (F-BIC) for fB-DNN classification:
\begin{equation}\label{EQ:FBIC}
    \textsc{F-BIC}\left(\bs{\eta}\right) = \sum_{i\in\mathcal{I}_2}\sum_{k=1}^K \mathbb{I}\left(Y_i=k\right)\left\{-\log \widehat{f}_{ik}\left(\bm X; \bs{\eta}\right) \right\} + {  C_{\tau} p_0\left(\bs{\eta}\right)} \frac{\log p}{n}, 
\end{equation}
where  the constant  $C_{\tau}=3\times 10^{\tau}$. In practice, we recommend $\tau = -1,0,1,2$. 
We name the associated algorithm ``fB-DNN".  The following Algorithms \ref{alg:selection} and \ref{alg:lassonet} describe the fB-DNN procedure, 
and Proposition \ref{proposition} validates the algorithms. The proof of 
Proposition \ref{proposition} is provided in Appendix. 
\begin{Proposition}\label{proposition}
 The solution to Algorithm \ref{alg:lassonet} is the global optimum of the constraint objective function (\ref{eq:obj}). 
\end{Proposition}
 {
Under Proposition \ref{proposition}, when the DNN architecture is properly selected,    the following theorem provides convergence rates  of our proposed classifier to the Bayes classifier. {Denote the true conditional density functions as $\left\{ f_{k} \right\}_{k=1}^K$. The orders $\alpha_1^\ast$ and $\alpha_2^\ast$ are defined in  the Appendix.}

{
\begin{theorem}\label{bayes}
   When the activation function $\sigma$ in (\ref{EQ:g}) is ReLU, it follows that: 
    \noindent{\it(a) } For $K=2$ and identity function $\sigma^\ast$ in (\ref{dnn class}), under Assumption 3 in the Appendix, if $f_1,f_2\in\mathcal{G}$ and the hinge loss is employed, 
     $\widehat{Y}$   defined in (\ref{DEF:Yhat}) satisfies  
    $$\Pro\left(\widehat{Y} \neq Y \right) - \Pro\left(\widehat{Y}^\ast \neq Y \right) \lesssim n^{-\alpha_1^\ast}\log^3n,$$
    where $\widehat{Y}^\ast$ is the class derived by the Bayes classifier of $\left(\bm X \right)$;
\noindent{\it(b) } For $K\geq 2$ and softmax function $\sigma^\ast$ in (\ref{dnn class}), under Assumption 4 in the Appendix, if $f_k\in\mathcal{G}_k$ and cross-entropy loss function (\ref{DEF:crossEn}) is employed,  the optimizer $\widehat{f}$ in (\ref{eq:obj})  satisfies  
$${\E}\left[\sum_{k=1}^K f_{k}\left(\bm X \right)\left(c_0\wedge \log\left(\frac{f_{k}\left(\bm X \right)}{\widehat {f}_k\left(\bm X \right)} \right) \right)\right]\lesssim n^{-\alpha_2^\ast}\log^3n,$$
where $c_0$ is a relatively large constant.
\end{theorem}}

 \begin{algorithm}
\SetKwInOut{Input}{Input}
\SetKwInOut{Output}{Output}
\Input{Functional data $\left\{\bm X_{i}\right\}_{i=1}^n$, class label $\left\{Y_i\right\}_{i=1}^n$, functional weights $\left\{\phi_{j,1},\phi_{j,2},\ldots, \right\}_{j=1}^p$, epoch number $B$, hierarchy coefficient $C$,  learning rate $\alpha$, 
 training index $\mathcal{I}_1$, validation index $\mathcal{I}_2$,  tuning candidates: $\left\{ \bm\eta_1, \ldots, \bm\eta_M\right\}$ with initial value $\lambda_0$ and path multiplier $\delta$, \;}
\For{$\ell=1,\ldots, M$}
{Apply Algorithm \ref{alg:lassonet} on $\left\{\bm X_i, Y_i \right\}_{i\in\mathcal{I}_1}$ and obtain $p_0\left(\bm \eta_\ell\right)$, $\bm b\left(\bm \eta_\ell\right)$ and $\bm W\left(\bm \eta_\ell\right)$\;
  Obtain the estimated probability $\widehat{f}_{ik}\left(\bm X;\bm \eta_\ell \right)$, $i\in\mathcal{I}_2$, $k=1,\ldots, K$ 
}
  Calculate the F-BIC for the validation dataset. Obtain $\bm \eta^\ast  = \underset{\bm\eta\in\left\{\bm \eta_1,\ldots, \bm\eta_M\right\}}{\argmin} \textsc{F-BIC}\left(\bm \eta \right)$\;
  Obtain the weights $\widehat{\bm b}\equiv{\bm b}(\bm\eta^\ast)$ and $\widehat{\bm W}\equiv{\bm W}(\bm\eta^\ast)$, and the selected index set $\widehat{\mathcal{A}} = \left\{j: \|\widehat{\bm b}_j \|\neq 0 \right\}$\;
\Output{$\widehat{\bm b}$, $\widehat{\bm W}$ and $\widehat{\mathcal{A}}$}
\caption{fB-DNN algorithm for functional data classification and feature selection }\label{alg:selection}
\end{algorithm}

\begin{algorithm}
\SetKwInOut{Input}{Input}
\SetKwInOut{Output}{Output}
\Input{Functional data $\left\{\bm X_{i}\right\}_{i=1}^n$, class label $\left\{Y_i\right\}_{i=1}^n$,  functional weights $\left\{\phi_{j,1},\ldots, \phi_{j,q_{1j}} \right\}$, number of hidden layer $L$, number of neurons $\bm q$, dropout rate $\gamma$, epoch number $B$, hierarchy coefficient $C$, path multiplier $\delta$, learning rate $\alpha$, initial tuning parameter $\lambda_0$.}
Initialize $\lambda=\lambda_0$, $p_0=p$ \;
\While{$p_0\geq 1$}
{ $\lambda \leftarrow (1+\delta)\lambda$
\For{$b=1,2,\ldots, B$}
{Compute $\nabla_{\bm b} \mathcal{L}$, $\nabla_{\bm W} \mathcal{L}$ by backpropagation \;
 Update ${\bm b}\leftarrow {\bm b} - \alpha \nabla_{\bm b} \mathcal{L}$, ${\bm W}\leftarrow {\bm W} - \alpha \nabla_{\bm W} \mathcal{L}$ \;
\For{$j=1,\ldots,p$}  
{Update $\bm b_j$ and $W_1^{(j)}$ via Algorithm \ref{Alg:W} in the Appendix.
 }
}}
 Update $p_0$ to be the number of non-zero coordinates of $\|\bm b_j\|$\;
\Output{$p_0$ and $\bm b$}
\caption{Training penalty term in Model (\ref{eq:obj})}\label{alg:lassonet}
\end{algorithm}

\section{Implementation}\label{sec:implement}
Our proposed neural networks are implemented using \texttt{Python} within the \texttt{PyTorch} framework. Specifically, the feedforward neural network is constructed based upon the number of hidden layer $L$, number of neurons in each hidden layer $q_l$, $l=0,\ldots, L$, and the dropout rate $\gamma$. Throughout, the epoch number and batch size for training the neural network are fixed before the training process. Practically, we choose the epoch number to be moderately large, as long as the results are stable, and the batch size equals $2^{\lfloor \log n\rfloor}$, where $\lfloor \cdot\rfloor$ denotes the floor function. We conduct Algorithm \ref{alg:selection} to choose the best neural network structure and the optimal truncation number. In practice, for any $j=1,\ldots, p$,  reasonable choices of functional weights expanding the $L^2\left(\left[0, 1\right]^d_j\right)$ yield comparable performance. This is because any misspecification merely induces a linear transformation, which is absorbed by the true function class $\mathcal{G}_k$ and therefore has a negligible effect. For all numerical studies, we apply Algorithm \ref{alg:selection} by randomly assigning $60\%$ data to the training set and $20\%$ data to the validation set, and the remaining $20\%$ data are used as a testing set to obtain the classification accuracy.
To optimize the training process with respect to objective (\ref{eq:obj}), we use \texttt{Adam} optimizer with a learning rate of $\alpha=0.001$ to train the initial model. The path multiplier is stable across all simulation settings, where we let $\delta=0.02$. We choose hierarchy coefficient $C=10$ as default. The optimization of Algorithm \ref{alg:lassonet} with respect to the  tuning parameter $\lambda$ adopts the dense-to-sparse warm start approach. This approach yields effective performance with strong generalization ability \citep{Lemhadri2018}.  { Python codes and examples for the proposed fB-DNN algorithms are available  on \texttt{GitHub}  (\url{https://github.com/ShuoyangWang/fB-DNN}).}

\section{Simulation studies}\label{sec:sim}
{In this section, we assess the effectiveness of our proposed method for handling 2D functional data and a mixture of 1D and 2D functional data. To gauge the comparative performance, { we consider four alternative methods, f-SVM, f-$k$NN, f-DNN and mfDNN \citep{wang:cao:23}. Support vector machine (SVM) and $k$-nearest neighbor ($k$NN) are  efficient classifiers designed for high-dimensional i.i.d. observations.  To make these methods directly applicable to functional data, we adapt them by utilizing  the inputs as functional principal scores extracted from all functional features. The realization is via the \texttt{R} packages \texttt{e1071} and \texttt{class}, respectively, where the tuning parameter candidates for the penalty term are generated by default. For f-SVM, we choose linear kernel and polynomial kernel with degree $3$. 
For f-$k$NN, $k$ is selected by cross-validation.} With f-DNN, we can demonstrate the comparative performance in selection and classification performance of our proposed F-BIC.  Note that the f-DNN procedure follows Algorithm \ref{alg:selection}, with a modification that substitutes the F-BIC criterion with the minimization of misclassification risk using  5-fold cross-validation. Specifically, in Algorithm \ref{alg:selection}, Step 6 is substituted with $\bm \eta^\ast  = \underset{\bm\eta\in\left\{\bm \eta_1,\ldots, \bm\eta_M\right\}}{\argmin} \textsc{MR}\left(\bm \eta \right)$, where  $\textsc{MR}\left(\bm \eta \right) = |\mathcal{I}_2|^{-1}\sum_{i\in\mathcal{I}_2} \mathbb{I}\left(Y_i\neq \widehat{Y}_i (\bm \eta) \right)$ is the empirical misclassification risk.}  To our knowledge, the mfDNN algorithm proposed by \cite{wang:cao:23} stands as the only publicly available algorithm tailored for both 1D and 2D functional data.  
Except for fB-DNN and f-DNN, the rest of competitors are not equipped with a regularization term to conduct feature selection.    Performance of selection is compared only between f-DNN and fB-DNN.  Results are based on  50 simulations.


\subsection{2D functional data case}\label{sec:sim2d}
We generated functional data $X_{ij}(s,t)= \sum_{\ell=1}^5 \xi_{ij,\ell} \phi_{\ell}(s,t) + \epsilon_{ij}(s,t)$, $i=1,\ldots,n_k$, $k=1,2,3$, $s,t\in\left[0, 1\right]$, where $\phi_1(s, t)=s$, $\phi_2(s, t)=t$, $\phi_3(s, t)=st$, $\phi_4(s, t)=s^2$, $\phi_5(s, t)=t^2$. The true feature set to distinguish the three classes is $\mathcal{A}=\left\{1,2,3,4,5 \right\}$.  {We generated the features in $\mathcal{A}^c = \left\{6,7,\ldots, p\right\}$ by letting 
$\bm\xi_{ij}|Y_i=k \sim \mathcal{N}\left((0,0,0,0,0)^\top, \right.$ $\left.\text{diag}(1,0.64,0.36,0.16,0.04)\right)$, $j=6,7,\ldots, p$,  where $\bm\xi_{ij} = \left(\xi_{ij,1},\xi_{ij,2}, \xi_{ij,3}\right)^\top$. } The features in $\mathcal{A}$ were generated by the following three settings:

\noindent {\bf Model I (Gaussian data with high separability):} 
     For $j=1,\ldots, 5$, let  \\
     $(\bm\xi_{ij}|Y_i=1) \sim $ $\mathcal{N}\left((2.5,2,1.5,1,0.5)^\top, \text{diag}(25,16,9,4,1)\right)$;\\
 $(\bm\xi_{ij}|Y_i=2) \sim \mathcal{N}\left((-2.5,-2,-1.5,-1,-0.5)^\top, \text{diag}(9,4,2.25,1,0.25)\right)$;\\
and  $(\bm\xi_{ij}|Y_i=3) \sim \mathcal{N}\left((0,0,0,0,0)^\top, \text{diag}(1,0.64,0.36,0.16,0.04)\right)$.

\noindent {\bf Model II (Gaussian data with low separability):}
For $j=1,\ldots, 5$, let \\
 $(\bm\xi_{ij}|Y_i=1) \sim \mathcal{N}\left((0.5,0.5,0.5,0.5,0.5)^\top,  \text{diag}(25,16,9,4,1)\right)$;  \\  $(\bm\xi_{ij}|Y_i=2) \sim \mathcal{N}\left((-0.5,-0.5,-0.5,-0.5,-0.5)^\top\text{diag}(9,4,2.25,1,0.25)\right)$;\\
    and $(\bm\xi_{ij}|Y_i=3) \sim \mathcal{N}\left((0,0,0,0,0)^\top, \text{diag}(1,0.64,0.36,0.16,0.04)\right)$.

\noindent {\bf Model III (Non-Gaussian data):}
For $j,\ell=1,\ldots, 5$, let \\
 $(\xi_{ij,\ell}|Y_i=1) \sim Exp(\theta_\ell), \text{ where }\theta_1=0.1, \theta_2=0.12, \theta_3=0.14, \theta_4=0.16, \theta_5=0.18$;\\
   $(\xi_{ij,\ell}|Y_i=2) \sim t_{2\ell+1}\left( 3\right)$;\\ 
  $(\bm\xi_{ij}|Y_i=3) \sim \mathcal{N}\left((0,0,0,0,0)^\top, \text{diag}(1,0.64,0.36,0.16,0.04)\right)$.
 
To mimic our real data of 2D PET scans, we consider functional features $p=50,100$, with sample size for each group $n_k \in \{100, 200\}$, $k=1,2,3$. Additionally, we chose $s$ and $t$ on $m$ evenly spaced grid points on $\left[0,1\right]$, with $m=30$ and $m=60$ {to represent relatively sparse and dense grid points}.
For each model, we observe the functional data on $30\times 30$ or $60\times 60$ grid points over $\left[ 0, 1\right]^2$.  The chosen optimal numbers of hidden layers and neurons in each layer vary by different replicates. The candidates for the number of hidden layers  $\{1,2,3\}$; and for the number of neurons in each layer $\left\{100, 300\right\}$.  
{ Table \ref{TAB:sim_gaussian1}  and \ref{TAB:merged_accuracy}  present the exact match rate (EMR) which is defined as the proportion that the selected  functional features exactly match the true functional features,  and the average  number of falsely selected features (FP). These two quantities describe the ability of various classifiers for feature selection. To measure the   classification accuracy, we report the average true classification rates.}

Several noteworthy findings emerge from the analysis of the two tables. Firstly, in Table \ref{TAB:sim_gaussian1}, both fB-DNN and f-DNN demonstrate superior performance in Models I and III compared to Model II. This discrepancy is expected given the low separability between classes in Model II, which makes  classification inherently more challenging. Secondly, fB-DNN exhibits an improved EMR with  increasing sample sizes, unlike f-DNN, where larger sample sizes lead to significantly worse performance in Model II. This observation suggests that f-DNN struggles to efficiently eliminate redundant information. 
{ This finding is unanimously supported by the notably high FP observed in f-DNN across all three models.} Lastly, in Table \ref{TAB:merged_accuracy},  fB-DNN and f-DNN display comparable classification accuracy and both  exhibit an enhanced accuracy with increasing sample sizes. On the other hand, { mfDNN and f-SVM alternative classifiers consistently lag behind, particularly in scenarios with a higher number of redundant features. f-$k$NN classifier has comparable classification rates in Models I and III. However, f-$k$NN has much lower classification accuracy in Model II.}
In summary, the simulation results underscore the superiority of the proposed fB-DNN method over alternative deep learning approaches when classifying 2D functional data. 


\begin{table}
\caption{\label{TAB:sim_gaussian1} 
Performance of feature selection for Models I, II, and III by the exact match rate, EMR (average number of incorrectly selected false features, FP).}
\centering
\begin{tabular}{@{\extracolsep{0.1pt}} ccccllll}
\hline
\hline   
\multirow{2}{*}{Model} &\multirow{2}{*}{Method} & \multirow{2}{*}{$n_k$} & \multicolumn{2}{c}{$m=30
$} & \multicolumn{2}{c}{$m=60$} \\   \cline{4-7}  
& &  &  $p=50$ & $p=100$ & \multicolumn{1}{l}{$p=50$} &  \multicolumn{1}{l}{$p=100$}  \\ \hline
\multirow{4}{*}{I} &\multirow{2}{*}{fB-DNN} &100 & \textbf{0.90(0.12)}  & \textbf{0.90(0.06)}  & \textbf{0.90(0.06)}  & \textbf{0.90(0.30)} \\ 
& &200 & \textbf{0.92(1.20)} & \textbf{0.92(0.12)} & \textbf{0.92(0.10)}  &  \textbf{0.92(0.06)} \\ 
&\multirow{2}{*}{f-DNN} &100 & 0.64(4.68)  & 0.40(14.96)  & 0.58(7.48)   & 0.54(10.50) \\ 
& &200 & 0.26(14.08)    & 0.20(24.30)  &  0.22(10.44)  & 0.18(28.18)\\  [3pt]  
\hline
\multirow{4}{*}{II} &\multirow{2}{*}{fB-DNN} &100 & \textbf{0.60(0.72)} & \textbf{0.46(0.94)}  & \textbf{0.72(0.50)}  & \textbf{0.70(0.56)} \\  
& &200 &    \textbf{0.68(2.16)} & \textbf{0.54(2.02)}  &  \textbf{0.76(1.10)} & \textbf{0.74(0.76)} \\ 
&\multirow{2}{*}{f-DNN} &100 & 0.26(11.42)  &  0.06(22.82) &  0.18(13.76)  & 0.24(18.66)  \\ 
& &200 & 0.08(31.66)    & 0.10(57.30)  & 0.08(28.10)   & 0.06(50.28)\\ [3pt]  
\hline
\multirow{4}{*}{III} &\multirow{2}{*}{fB-DNN} &100 & \textbf{0.98(0.04)}  & \textbf{0.96(0.04)}  & \textbf{0.98(0.04)}  & \textbf{0.96(0.12)} \\ 
& &200 &  \textbf{1.00(0.00)} & \textbf{1.00(0.00)} &  \textbf{1.00(0.00)} &   \textbf{1.00(0.00)} \\  
&\multirow{2}{*}{f-DNN} &100 & 0.84(3.44)   &  0.80(1.38)  &  0.76(3.06)  & 0.84(0.92) \\   
& &200 &  0.90(2.46)    & 0.98(2.72)  & 0.88(2.26)   & 0.96(2.86)\\  [3pt]   
\hline\hline
\end{tabular}
\end{table}


\begin{table}
\caption{\label{TAB:merged_accuracy}  The average true classification rates  for Models I, II, and III.}
\centering
\begin{tabular}{@{\extracolsep{0.1pt}} cccccccc}
\hline
\hline   
\multirow{2}{*}{Model} &\multirow{2}{*}{Method} & \multirow{2}{*}{$n_k$} & \multicolumn{2}{c}{$m=30
$} & \multicolumn{2}{c}{$m=60$} \\   \cline{4-7}  
& &  & \multicolumn{1}{c}{$p=50$} &  \multicolumn{1}{c}{$p=100$} & \multicolumn{1}{c}{$p=50$} &  \multicolumn{1}{c}{$p=100$}  \\ \hline
& \multirow{2}{*}{fB-DNN} &100 & \textbf{0.974}  & \textbf{0.964} & \textbf{0.970}  & 0.969  \\  
& &200 & 0.980  & \textbf{0.982}  & 0.981  & \textbf{0.983}   \\  
&\multirow{2}{*}{f-DNN} &100 & 0.964  & 0.964 & 0.968  & \textbf{0.970} \\  
& &200 & \textbf{0.982}  & 0.980  & \textbf{0.983}  & 0.981 \\  
\multirow{2}{*}{I}&\multirow{2}{*}{mfDNN} &100 & 0.935  & 0.296 & 0.936  & 0.298 \\  
& &200 & 0.957  & 0.303  & 0.954  & 0.304 \\  
&\multirow{2}{*}{f-SVM} &100 & 0.896  & 0.817 & 0.895  & 0.821 \\  
& &200 & 0.928  & 0.894  & 0.928  & 0.897 \\  
&\multirow{2}{*}{f-$k$NN} &100 & 0.909  & 0.906 & 0.909  & 0.906 \\  
& &200 & 0.933  & 0.931  & 0.933  & 0.931 \\  
 [3pt]\hline
&\multirow{2}{*}{fB-DNN} &100 & \textbf{0.854}  & \textbf{0.850} & 0.846  & \textbf{0.857} \\  
& &200 & \textbf{0.872}  & \textbf{0.871}  & \textbf{0.869}  & \textbf{0.871} \\  
&\multirow{2}{*}{f-DNN} &100 & 0.854  & 0.845 & \textbf{0.850}  & 0.850 \\  
& &200 & 0.866  & 0.869  & 0.864  & 0.866 \\  
\multirow{2}{*}{II}&\multirow{2}{*}{mfDNN} &100 & 0.677  & 0.292 & 0.669  & 0.300 \\  
& &200 & 0.751  & 0.301  & 0.751  & 0.306 \\  
&\multirow{2}{*}{f-SVM} &100 & 0.579  & 0.508 & 0.586 & 0.511 \\  
& &200 & 0.647  & 0.556  & 0.654  & 0.564 \\  
&\multirow{2}{*}{f-$k$NN} &100 & 0.588  & 0.593 & 0.587  & 0.594 \\  
& &200 & 0.667  & 0.643  & 0.667  & 0.643 \\  
 [3pt]\hline
&\multirow{2}{*}{fB-DNN} &100 & \textbf{0.964}  & \textbf{0.961} & \textbf{0.964}  & 0.957 \\  
& &200 & \textbf{0.966}  & \textbf{0.968}  & \textbf{0.978}  & \textbf{0.961} \\  
&\multirow{2}{*}{f-DNN} &100 & 0.945  & 0.952 & 0.953  & \textbf{0.959} \\  
& &200 & 0.959  & 0.961  & 0.961  & 0.955 \\  
\multirow{2}{*}{III}&\multirow{2}{*}{mfDNN} &100 & 0.872  & 0.302 & 0.876  & 0.297 \\  
& &200 & 0.934  & 0.312  & 0.931  & 0.315 \\
&\multirow{2}{*}{f-SVM} &100 & 0.933  & 0.867 & 0.930  & 0.865 \\  
& &200 & 0.959  & 0.939  & 0.957  & 0.937 \\  
&\multirow{2}{*}{f-$k$NN} &100 & 0.920  & 0.923 & 0.921  & 0.923 \\  
& &200 & 0.932  & 0.934  & 0.932  & 0.934 \\  
 [3pt]\hline
\hline
\end{tabular}
\end{table}

\subsection{Mixture of 1D and 2D functional data case}\label{sec:sim1d2d}
We first generated 1D functional data $X_{ij}(s)= \sum_{\ell=1}^3 \xi_{ij,\ell} \phi_{\ell}(s) + \epsilon_{ij}(s)$, $i=1,\ldots,n_k$, $k=1,2,3$, $s\in\left[0, 1\right]$, where $\phi_1(s)=\log(s+2)$, $\phi_2(s)=s$, $\phi_3(s)=s^3$. Across all settings, we consider $30$ features of 1D data to mimic our real dataset, where we let the true feature be $\mathcal{A}_1=\left\{1,2,3\right\}$. We generated the features in $\mathcal{A}_1^c = \left\{4,5,\ldots, 30\right\}$ by letting 
$\bm\xi_{ij}|Y_i=k \sim \mathcal{N}\left((0,0,0)^\top, \text{diag}(1,0.64,0.36)\right)$, $j=4,5,\ldots, 30$, where $\bm\xi_{ij} = \left(\xi_{ij,1},\xi_{ij,2}, \xi_{ij,3}\right)^\top$. For 2D functional data, we kept the same true feature set $\mathcal{A}_2=\left\{1,\ldots, 5\right\}$ whose specifications followed the same details in Section \ref{sec:sim2d}. The features in $\mathcal{A}_1$ and $\mathcal{A}_2$ were generated by the following three settings:

\noindent {\bf Model IV (Gaussian data with high separability)}
For 1D functional data with $j=1,2,3$, let \\
  $(\bm\xi_{ij}|Y_i=1) \sim \mathcal{N}\left((2.5,2,1.5)^\top, \text{diag}(25,16,9)\right)$;\\
 $(\bm\xi_{ij}|Y_i=2) \sim \mathcal{N}\left((-2.5,-2,-1.5)^\top, \text{diag}(9,4,2.25)\right)$;\\
  $(\bm\xi_{ij}|Y_i=3) \sim \mathcal{N}\left((0,0,0)^\top, \text{diag}(1,0.64,0.36)\right)$.
2D functional data are generated as Model I.

\noindent {\bf Model V (Gaussian data with low separability):}
For 1D functional data with $j=1,2,3$, let \\
$(\bm\xi_{ij}|Y_i=1) \sim \mathcal{N}\left((0.5,0.5,0.5)^\top, \text{diag}(25,16,9)\right)$;\\
$(\bm\xi_{ij}|Y_i=2) \sim \mathcal{N}\left((-0.5,-0.5,-0.5)^\top, \text{diag}(9,4,2.25)\right)$;\\
$(\bm\xi_{ij}|Y_i=3) \sim \mathcal{N}\left((0,0,0)^\top, \text{diag}(1,0.64,0.36)\right)$.
2D functional data are generated as Model II.

\noindent {\bf Model VI (Non-Gaussian data):}
For 1D functional data with $j=1,2,3$, let \\
  $(\xi_{ij,\ell}|Y_i=1) \sim Exp(\theta_\ell), \text{ where }$ $\theta_1=0.1$, $\theta_2=0.15$, $\theta_3=0.2$;\\
  $(\xi_{ij,\ell}|Y_i=2) \sim t_{2\ell+2}\left( 3\right)$, $\ell=1,2,3$;\\
 $(\bm\xi_{ij}|Y_i=3) \sim \mathcal{N}\left((0,0,0)^\top, \text{diag}(1.2,0.8,0.4)\right)$.
  2D functional data are generated as Model III.

\begin{table}
\caption{\label{TAB:sim_gaussian1_mixed} Performance of feature selection for Models IV, V, and VI by the exact match rate, EMR (average number of incorrectly selected false features, FP).}
\centering
\begin{tabular}{@{\extracolsep{0.1pt}} ccccllll}
\hline
\hline   
\multirow{2}{*}{Model}&\multirow{2}{*}{Method} & \multirow{2}{*}{$n_k$} & \multicolumn{2}{c}{$m=30$} & \multicolumn{2}{c}{$m=60$} \\   \cline{4-7}  
&&  & \multicolumn{1}{c}{$p=80$} &  \multicolumn{1}{c}{$p=130$} & \multicolumn{1}{c}{$p=80$} &  \multicolumn{1}{c}{$p=130$}  \\ \hline
\multirow{4}{*}{IV}&fB-DNN &100 & \textbf{0.88(0.14)}  & \textbf{0.74(0.32)} & \textbf{0.88(0.10)}  & \textbf{0.82(0.28)} \\  [1pt] 
& &200 & \textbf{0.86(0.28)} & \textbf{0.86(0.38)} & \textbf{0.86(0.18)}  &  \textbf{0.84(0.34)} \\  [1pt] 
&f-DNN&100 & 0.26(10.04) & 0.14(16.52) & 0.30(7.38)  &  0.20(18.06) \\  [1pt] 
& &200 & 0.06(36.96)  & 0.04(64.58) &  0.04(38.00) &  0.02(57.32) \\  [3pt] \hline
\multirow{4}{*}{V}&fB-DNN &100 & \textbf{0.84(0.08)} & \textbf{0.86(0.06)} & \textbf{0.86(0.02)}  & \textbf{0.80(0.06)} \\  [1pt] 
& &200 & \textbf{0.84(0.06)}  & \textbf{0.84(0.06)} & \textbf{0.86(0.00)}    & \textbf{0.84(0.04)}  \\  [1pt] 
&f-DNN&100 & 0.70(7.98)  & 0.62(14.52) & 0.68(7.04)  & 0.58(16.96) \\  [1pt] 
& &200 & 0.28(14.98)  & 0.30(24.58) & 0.32(12.72)  &  0.26(19.08) \\ [3pt]\hline
\multirow{4}{*}{VI}&fB-DNN &100 & \textbf{0.66(0.16)}  & \textbf{0.60(0.36)} & \textbf{0.62(0.38)}  & \textbf{0.62(0.66)} \\  [1pt]
& &200 & \textbf{0.82(0.20)}  &  \textbf{0.80(0.48)} &  \textbf{0.84(0.10)} &  \textbf{0.88(0.00)} \\  [1pt] 
&f-DNN&100 & 0.56(1.26)  & 0.46(3.42) & 0.40(2.90)  & 0.48(3.50) \\  [1pt] 
& &200 & 0.76(7.98)  & 0.52(2.10) &  0.80(6.54) &  0.80(3.18)  \\ [3pt] \hline
\hline
\end{tabular}
\end{table}

 \begin{table}
\caption{\label{TAB:sim_gaussian1_acc_mixed} The average true classification rates  for Models IV, V, and VI.}
\centering
\begin{tabular}{@{\extracolsep{0.1pt}} cccccccc}
\hline
\hline   
\multirow{2}{*}{Model} &\multirow{2}{*}{Method} & \multirow{2}{*}{$n_k$} & \multicolumn{2}{c}{$m=30$} & \multicolumn{2}{c}{$m=60$} \\   \cline{4-7}  
&&  & \multicolumn{1}{c}{$p=80$} &  \multicolumn{1}{c}{$p=130$} & \multicolumn{1}{c}{$p=80$} &  \multicolumn{1}{c}{$p=130$}  \\ \hline
&\multirow{2}{*}{fB-DNN} &100 & \textbf{0.876} & \textbf{0.878} & \textbf{0.881} & \textbf{0.879} \\  
& &200 & 0.912 & \textbf{0.912} & 0.911 & 0.911 \\ 
&\multirow{2}{*}{f-DNN} &100 & 0.881 & 0.877 & 0.880 & 0.879 \\ 
& &200 & \textbf{0.913} & 0.910 & \textbf{0.912} & \textbf{0.912} \\  
\multirow{2}{*}{IV}&\multirow{2}{*}{mfDNN} &100 & 0.302 & 0.301 & 0.290 & 0.301 \\  
& &200 & 0.308 & 0.303 & 0.313 & 0.303 \\  
&\multirow{2}{*}{f-SVM} &100 &0.397 & 0.370 & 0.390 & 0.364 \\   
& &200 & 0.573 & 0.475 & 0.560 & 0.464 \\ 
&\multirow{2}{*}{f-$k$NN} &100 &0.867 & 0.868 & 0.867 & 0.868 \\   
& &200 & 0.911 & 0.903 & 0.911 & 0.903 \\ [3pt] \hline
&\multirow{2}{*}{fB-DNN} &100 & 0.970 &  0.970& 0.977 & \textbf{0.978} \\  
& &200 & \textbf{0.987} &  \textbf{0.986}& \textbf{0.989} & 0.979 \\ 
&\multirow{2}{*}{f-DNN} &100 & \textbf{0.973} & \textbf{0.974} & \textbf{0.978} & 0.973 \\  
& &200 & 0.985 & 0.986  & 0.985 &  \textbf{0.986}\\  
\multirow{2}{*}{V}&\multirow{2}{*}{mfDNN} &100 &0.304 & 0.294 & 0.300 & 0.292 \\   
& &200 & 0.306 & 0.305 & 0.310 & 0.315 \\ 
&\multirow{2}{*}{f-SVM} &100 &0.319 & 0.315 & 0.319 & 0.316 \\   
& &200 & 0.356 & 0.342 & 0.354 & 0.340 \\ 
&\multirow{2}{*}{f-$k$NN} &100 &0.485 & 0.466 & 0.485 & 0.466 \\   
& &200 & 0.584 & 0.565 & 0.582 & 0.564 \\ [3pt]
 \hline
&\multirow{2}{*}{fB-DNN} &100 & \textbf{0.963} & \textbf{0.961} & \textbf{0.964} & \textbf{0.961} \\ 
& &200 & \textbf{0.966} & \textbf{0.968} & \textbf{0.978} & \textbf{0.966} \\ 
&\multirow{2}{*}{f-DNN} &100 & 0.961 & 0.932 & 0.959 & 0.946 \\  
& &200 & 0.961 & 0.948 & 0.967 & 0.965  \\ 
\multirow{2}{*}{VI}& \multirow{2}{*}{mfDNN} &100 & 0.291 & 0.301 & 0.302 & 0.289 \\  
& &200 & 0.310 & 0.315 & 0.308 & 0.310 \\ 
&\multirow{2}{*}{f-SVM} &100 &0.458 & 0.402 & 0.451 & 0.392 \\   
& &200 & 0.651 & 0.485 & 0.637 & 0.467 \\ 
&\multirow{2}{*}{f-$k$NN} &100 &0.957 & 0.954 & 0.958 & 0.954 \\   
& &200 & 0.964 & 0.962 & 0.964 & 0.961 \\ [3pt] 
\hline
\hline
\end{tabular}
\end{table}

We chose to observe the 1D functional data on $15$ evenly spaced grid points on $\left[0,1\right]$. The number of features for 2D functional data, number of grid points $m$, and sample sizes $n_k$ were chose the same as described in Section \ref{sec:sim2d}.  The chosen optimal numbers of hidden layers and neurons in each layer  also followed the same setting in  in Section \ref{sec:sim2d}.
For each model,   the number of total features  $p \in \{80,130\}$, accounting for both the 1D and 2D functional features. Table \ref{TAB:sim_gaussian1_mixed} shows the EMR and FP for Models IV to VI.  Table \ref{TAB:sim_gaussian1_acc_mixed} depicts their classification accuracy.

The proposed fB-DNN classifier consistently outperforms its counterparts across all mixture functional data scenarios. 
Specifically, in Model IV, when $n=100$, the EMR of f-DNN is approximately one-fourth to one-third of fB-DNN's EMR; when $n=200$, the EMR of f-DNN is only one-twentieth of fB-DNN's EMR.   Notably, fB-DNN's FP is  extremely smaller compared to that of f-DNN across all  settings.   This discrepancy suggests that f-DNN struggles to correctly discern important features. In contrast, our proposed fB-DNN efficiently extracts significant features from the continuum within the  mixture functional data framework. Furthermore, fB-DNN demonstrates the ability to uncover underlying distributions of functional data clusters effectively.
Similar to the 2D functional data cases, both f-DNN and fB-DNN exhibit comparable performance across all scenarios, {while the classification accuracy of alternative methods is consistently one-half or less than one-third of fB-DNN's accuracy in Model V, suggesting the possible detrimental impact of retaining redundant features on classification performance.} Once again, our proposed classifier demonstrates a distinct advantage over its competitors in addressing multivariate functional data classification challenges.

\section{Application for Alzheimer's Disease}\label{SEC:realdata}
Brain scans, which can be considered as 2D or 3D functional data, play a pivotal role in  the diagnosis of  AD. Accurately identifying the affected regions while simultaneously excluding areas without pathology is crucial. This not only facilitates the development of targeted treatment strategies but also helps spare individuals from unnecessary procedures or therapies, ensuring that resources are aptly allocated. As an application, we used the dataset obtained from the ADNI database (\url{http://adni.loni.usc.edu}). The ADNI is a longitudinal multicenter study designed to develop clinical, imaging, genetic, and biochemical biomarkers for the early detection and tracking disease progression of  AD. From this database, we included $79$ patients in AD group, $45$ patients in  {EMCI} group, and $101$ people in  {CN} group whose PET data are available. 

For AD group, patients' age ranges from $59$ to $88$, with an   average age of $76.49$; and there are $33$ females and $46$ males among these $79$ subjects. For EMCI group, patients' age ranges from $57$ to $89$, with an average age of  $72.33$; and there are $26$ females and $19$ males among these $45$ subjects. For {CN} group, patients' age ranges from $62$ to $87$, with an  average age of $75.98$; and there are $40$ females and $61$ males among these $101$ subjects.  We collected $10$ 1D functional data for each subject. These variables represent time-based biomarkers, recorded from the start of each patient's participation in the study, and the numbers of observations for each variable vary from $2$ to $15$. The variable names and their descriptions are listed in Table \ref{tab:ad_biomarkers}. {Figure \ref{FIG:1Drealdata} depicts four functional biomarkers observed varying over $180$ months.} All 2D PET scans were reoriented into $79\times 95 \times 68$ voxels, which means that each patient has $68$ sliced 2D images with  $79\times 95$ pixels.  {This  PET  dataset has been  post-processed. }

\begin{table}[H]
\centering
\caption{Summary of Functional Biomarkers for Alzheimer's Disease}
\label{tab:ad_biomarkers}
\begin{tabular}{|l|l|}
\hline
Variable Name & Description \\
\hline
1. FDG & Fluorodeoxyglucose-PET measures for cerebral metabolic rates \\
2. CDRSB & Clinical Dementia Rating-Sum of Boxes \\
3. ADAS13 & AD Assessment Scale-Cognitive Subscale (13 items) \\
4. ADASQ4 & Delayed Word Recall Score \\
5. MMSE & Mini-Mental State Examination \\
6. RAVLT.learning & Rey Auditory Verbal Learning Test (learning score) \\
7. RAVLT.forgetting & Rey Auditory Verbal Learning Test (forgetting score) \\
8. LDELTOTAL & Logical Memory Delayed Recall (total score) \\
9. mPACCdigit & Modified Preclinical Alzheimer Cognitive Composite \\
&(digit symbol-coding) \\
10. mPACCtrailsB & Modified Preclinical Alzheimer Cognitive Composite\\
&(Trail Making Test Part B) \\
\hline
\end{tabular}
\end{table}

\begin{figure}
    \centering
        $\begin{array}{l}
\includegraphics[width = 0.45\textwidth]{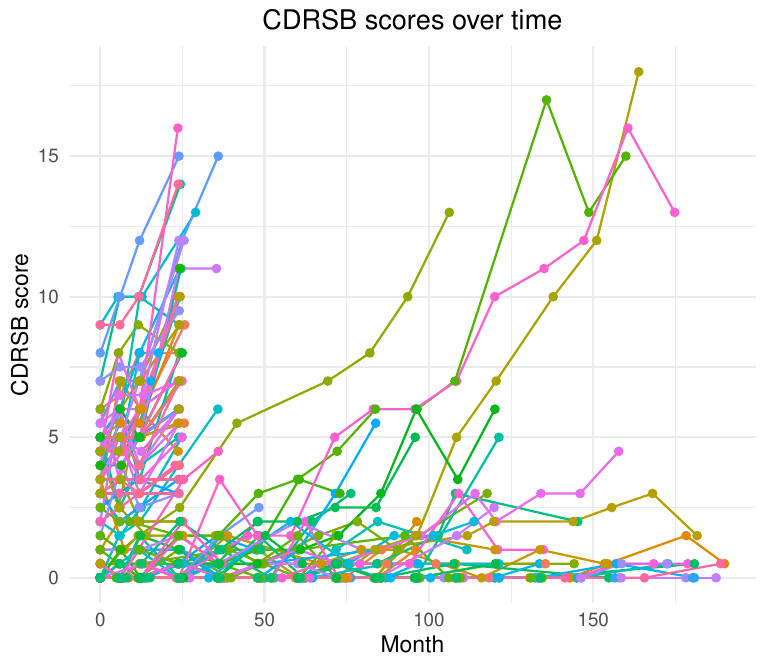} 
\hspace{1.mm}
\includegraphics[width = 0.45\textwidth]{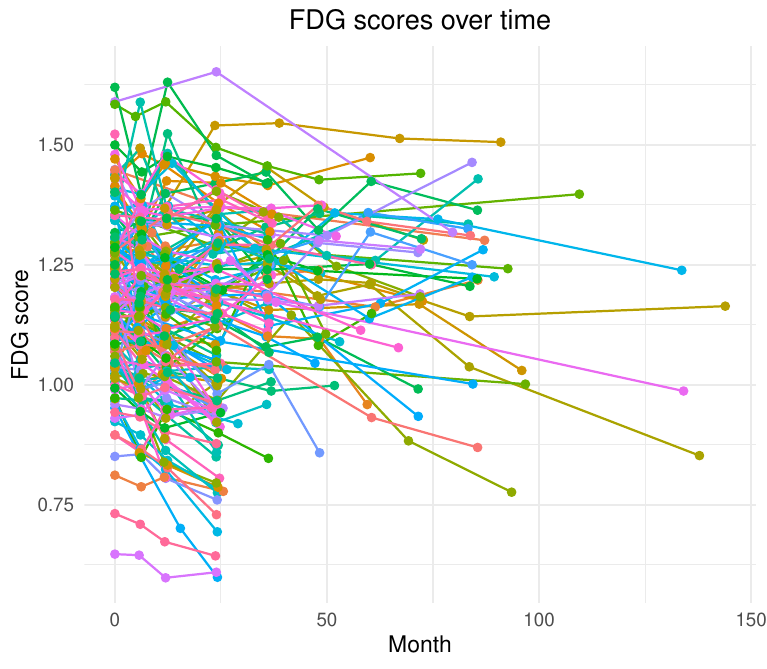} \\
\end{array}$\\
$\begin{array}{l}
\includegraphics[width = 0.45\textwidth]{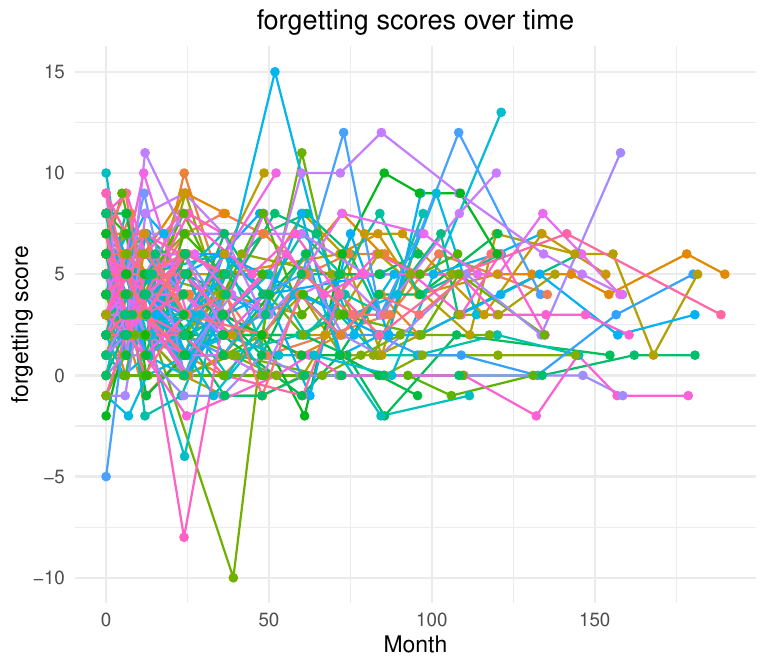} 
 \hspace{1.mm}
 \includegraphics[width = 0.45\textwidth]{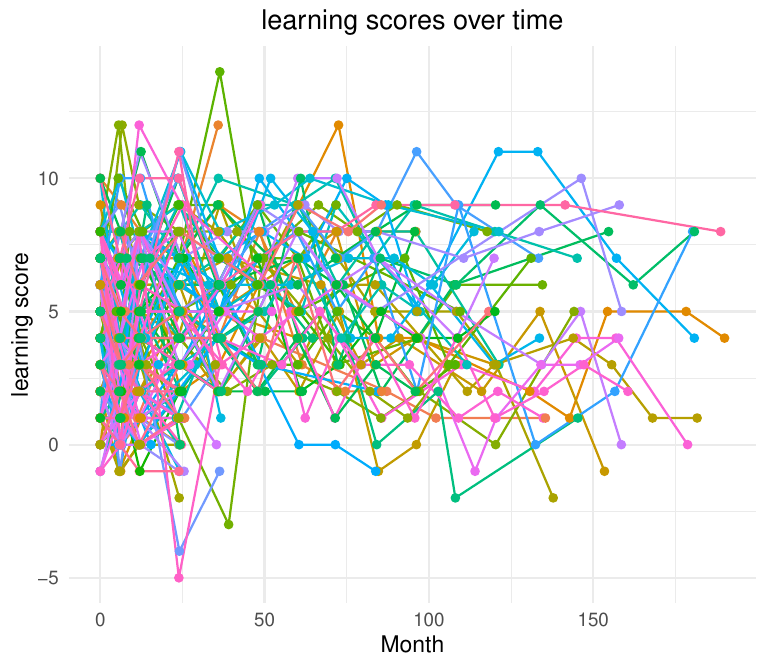} \\
\end{array}$
  \caption{Four functional biomarkers from ADNI dataset for each subject. Top: Clinical Dementia Rating-Sum of Boxes (left) and Fluorodeoxyglucose-PET measures for cerebral metabolic rates (right). Bottom: Rey Auditory Verbal Learning Test learning score (left) and forgetting score (right).}
    \label{FIG:1Drealdata}
\end{figure}

\begin{figure}
\begin{center}
\hspace{.4cm}\textbf{AD} \hspace{2.8cm}\textbf{EMCI}\hspace{2.6cm}\textbf{CN}\\
$20$-th
$\begin{array}{l}
\includegraphics[width = 0.28\textwidth]{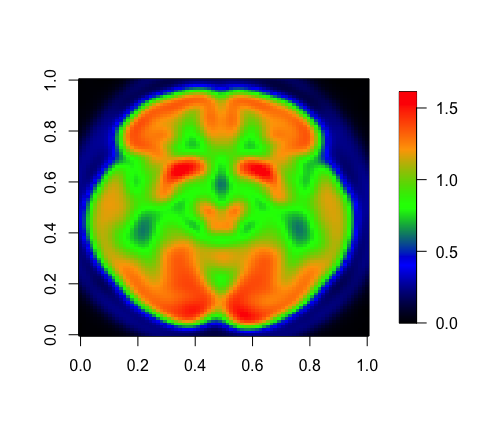} 
\hspace{1.mm}
\includegraphics[width = 0.28\textwidth]{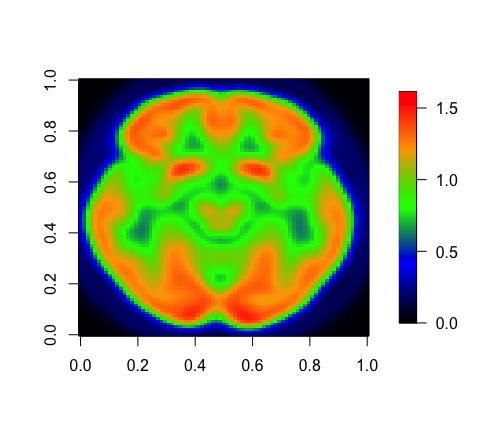} 
\hspace{1.mm}
\includegraphics[width = 0.28\textwidth]{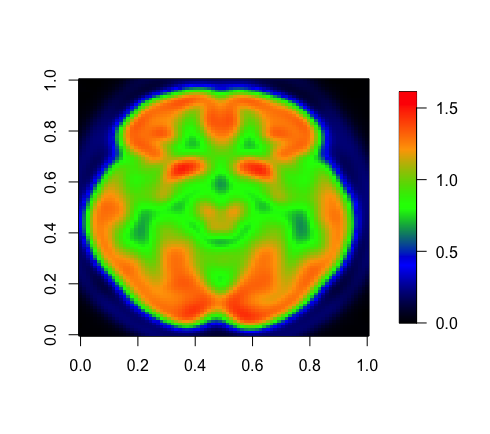} \\
\end{array}$\\
\vspace{-.6cm}
$40$-th
$\begin{array}{l}
\includegraphics[width = 0.28\textwidth]{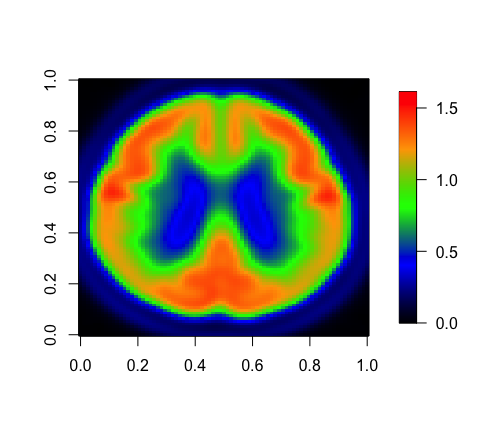} 
\hspace{1.mm}
 \includegraphics[width = 0.28\textwidth]{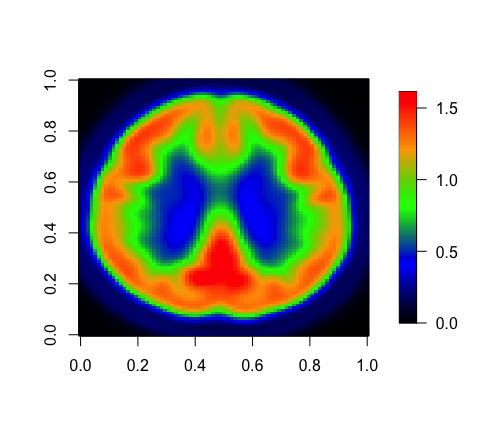} 
 \hspace{1.mm}
 \includegraphics[width = 0.28\textwidth]{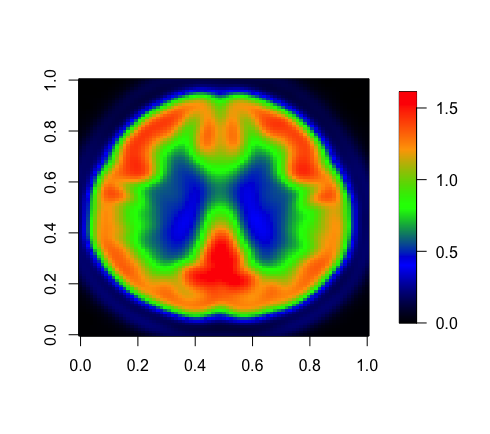} \\
\end{array}$\\
\vspace{-.6cm}
$60$-th
$\begin{array}{l}
\includegraphics[width = 0.28\textwidth]{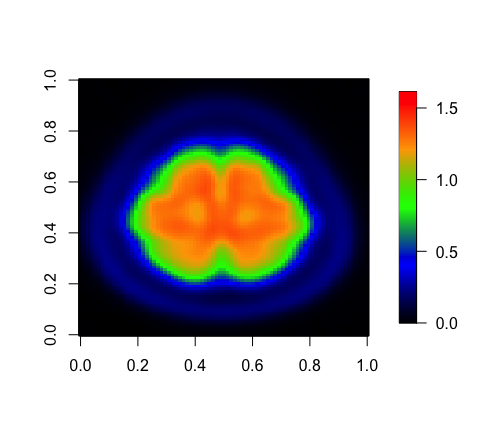} 
\hspace{1.mm}
 \includegraphics[width = 0.28\textwidth]{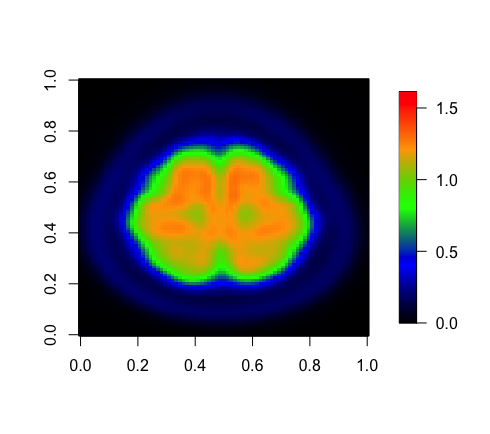} 
 \hspace{1.mm}
 \includegraphics[width = 0.28\textwidth]{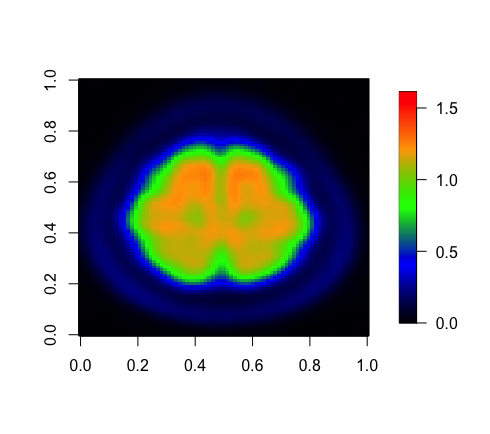} \\
\end{array}$\\
\vspace{-.6cm}
\caption{\label{fig:ADNI_2d} The $20$-th,  the $40$-th and the $60$-th slices of AD  group (left column), EMCI group (middle column), and CN  group (right column) from three random samples.} 
\end{center}
\end{figure}

Given the complex and voluminous nature of the functional data, our objective is twofold: to provide physicians with insights into critical disease features and to construct a high-performance classifier for diagnosing the status of new patients. 
{
Implementation follows details in Section \ref{sec:implement}.  Regarding input data, we consider classification based sole on  the 2D imaging data, as well as datasets comprising a mixture of 1D functional data (time-based biomarkers) and 2D functional data (2D images). The classification and feature selection results are summarized in Table \ref{TAB:real data}.} 
\begin{table}
\caption{\label{TAB:real data} Classification Accuracy and Selected Features of Different Models for ADNI 
Dataset.}
\centering
\begin{tabular}{@{\extracolsep{0.1pt}} cccccccc}
\hline
\hline   
Data&{Model} & {Accuracy (\%)} & \multicolumn{4}{c}{Selected features} \\   \hline
& mfDNN & 52.9\%  & \multicolumn{4}{c}{---}\\ [2pt]
& f-SVM & 50.0\%  & \multicolumn{4}{c}{---}\\ [2pt]
{2D} & f-$k$NN & 51.5\%  & \multicolumn{4}{c}{---}\\ [2pt]
& f-DNN& 61.8\%  & \multicolumn{4}{c}{Scans 16--29}  \\ [2pt]
& fB-DNN  & \textbf{63.2\%} & \multicolumn{4}{c}{Scans 18--25}\\
\hline 
&mfDNN & 61.8\% & \multicolumn{4}{c}{---}\\ [2pt]
&f-SVM & 51.5\% & \multicolumn{4}{c}{---}\\ [2pt]
{1D \& 2D} &f-$k$NN & 58.8\% & \multicolumn{4}{c}{---}\\ [2pt]
 & f-DNN& 63.2\%  & \multicolumn{4}{c}{Biomarkers 2--10, Scan 61} \\
& fB-DNN & \textbf{70.6\%} & \multicolumn{4}{c}{Biomarkers 2--9, Scans 15--35} \\
\hline
\hline
\end{tabular}
\end{table}

{ Clearly, all classifiers have improved classification accuracy when 1D functional data is included along with 2D functional data. This observation underscores the importance of leveraging multivariate multidimensional functional data. Both fB-DNN and f-DNN exhibit higher classification accuracy compared to alternative methods.
When comparing fB-DNN and f-DNN, fB-DNN shows slightly better performance. However, f-DNN selects a questionable slice (\#61) in the 1D \& 2D setting. It is well known that Alzheimer's disease destroys neurons and their connections in the hippocampus, the entorhinal cortex, and the cerebral cortex. These regions correspond to the first $25$ slices, as discussed in the real data analysis results by \cite{wang:etal:23}.
Notably, our proposed fB-DNN algorithm effectively selects reasonable and significant features, presenting promising findings for neurologists. The 2D slices and biomarkers identified by the algorithm offer valuable insights for distinguishing between CN, EMCI, and AD groups, guiding further medical evaluations targeted at these critical brain locations and biomarkers. Therefore, considering both classification and feature selection accuracy, fB-DNN stands out among existing classifiers.
}

 \vskip -3cm
 \section{Discussion}\label{SEC:summary}

 This paper introduces a novel method for performing classification of multivariate   functional data using DNNs framework. The proposed fB-DNN algorithm establishes a relationship between a scalar outcome and a set of predictors through a DNN-based approach, enabling it to handle various factors and correlations between functions effectively. Additionally, it efficiently selects significant functional variables, thereby enhancing classification accuracy. Leveraging the principles of Lasso regularization, the method effectively manages large numbers of functional covariates and image data. Moreover, it demonstrates scalability to large datasets through automated coding, with potential for further flexibility in accommodating mixed curved data and image data. Notably, this work represents the first unified framework for multiclassification and feature selection in the context of multivariate multidimensional functional data.

 To the best of our knowledge, even for simple i.i.d. data, no existing work has addressed feature selection consistency for DNN that incorporate residual layers and penalty functions. This represents a challenging and promising direction for our future research in functional data classification.

\section*{Acknowledgment}

Data used in preparation of this article were obtained from the Alzheimer’s Disease Neuroimaging Initiative (ADNI) database (\url{http://adni.loni.usc.edu}). As such, the investigators within the ADNI contributed to the design and implementation of ADNI and/or provided data but did not participate in analysis or writing of this report. A complete listing of ADNI investigators can be found at: \url{http://adni.loni.usc.edu/wp-content/uploads/how_to_apply/ADNI_Acknowledgement_List.pdf}.

 
\section*{Appendix}
\newtheorem{proposition}[theorem]{Proposition}
\newtheorem{corollary}[theorem]{Corollary}

\renewcommand{\thetheorem}{{\sc A.\arabic{theorem}}}
\renewcommand{\theproposition}{{\sc A.\arabic{proposition}}}
\renewcommand{\thelemma}{{\sc A.\arabic{lemma}}}
\renewcommand{\thecorollary}{{\sc A.\arabic{corollary}}} %
\renewcommand{\theequation}{A.\arabic{equation}}
\renewcommand{\thesubsection}{{\it A.\arabic{subsection}}} \setcounter{equation}{0} %
\setcounter{lemma}{0} 
\setcounter{theorem}{0} %
\setcounter{subsection}{0}

In this Appendix, we provide additional numerical results and Algorithm \ref{Alg:W}. Technical assumptions and proofs are available from the authors upon request. 
\subsection{Computational time}
The computation job of this paper is conducted at Michigan State University high performance computing center,  where each compute node containing multiple CPU cores and substantial memory (RAM). The specific hardware configurations may vary, but it always includes  {$1$ CPU and $12$ GB RAM as default for each replicates} of the parallel computing. 

 The computing times for different models in the simulation settings are very similar to each other.  
Table \ref{TAB:time} presents the average computational time (in minutes) for Model I. As expected, all DNN-based methods are computationally more expensive than the f-SVM and f-$k$NN methods. This can be attributed to several factors, including layer depth, model complexity, parameter volume, and memory requirements, as discussed in classic deep learning literature \cite{lecun2015deep}. Reducing the computational cost of the proposed fb-DNN method represents a promising direction for future research. Since fb-DNN performs both feature selection and classification simultaneously, it requires additional computational resources compared to f-DNN and mfDNN, which do not include a feature selection process.
\begin{table}
\caption{\label{TAB:time}  The average computational time (in minutes)  for Model I.}
\centering
\begin{tabular}{@{\extracolsep{0.1pt}} ccccccc}
\hline
\hline   
 \multirow{2}{*}{Method} & \multirow{2}{*}{$n_k$} & \multicolumn{2}{c}{$m=30
$} & \multicolumn{2}{c}{$m=60$} \\   \cline{3-6}  
 &  & \multicolumn{1}{c}{$p=50$} &  \multicolumn{1}{c}{$p=100$} & \multicolumn{1}{c}{$p=50$} &  \multicolumn{1}{c}{$p=100$}  \\ \hline
 \multirow{2}{*}{fB-DNN} &100 & 280.18   & 348.87 & 245.04  & 374.39 \\  
 &200  &  290.74  & 432.24 &  304.46  & 382.51 \\   
\multirow{2}{*}{f-DNN} &100 & 90.02  & 94.33 & 90.83   & 91.79 \\  
 &200 & 87.46   & 100.62 & 89.88   & 95.39 \\   
\multirow{2}{*}{mfDNN} &100  & 24.63   & 25.04 & 24.68   &  25.35\\    
 &200  & 25.46   & 25.98 & 25.62   & 26.14 \\   
\multirow{2}{*}{f-SVM} &100 & 0.14   & 0.15 & 0.15   &  0.15\\  
 &200 &  0.17  & 0.19 & 0.17   & 0.18 \\  
\multirow{2}{*}{f-$k$NN} &100 &  0.13  & 0.14 &  0.12  &  0.14\\  
 &200  & 0.16   & 0.18 & 0.14   & 0.19 \\  
 [3pt]\hline
\hline
\end{tabular}
\end{table}


\subsection{Algorithm 3}
\begin{algorithm}[H] 

\SetKwInOut{Input}{Input}
\SetKwInOut{Output}{Output}
\Input{hierarchy coefficient $C$, learning rate $\alpha$,   network weights ${\bm b}_j$  and ${W_j}$.}
  Sort $\text{vec}\left(W_1^{(j)}\right)\in\mathbb{R}^{\nu_j}$ and denote by $|\omega_{j(1)}|\geq \ldots \geq |\omega_{j(\nu_j)}|$, where $\nu_j=q_{1j}q_2$\;
\For{$u=0,\ldots,\nu_j$} 
{ Calculate $\zeta_{ju} = \frac{C}{1+C^2}{S}_{\alpha\lambda}\left(\|\bm b_j\| + C\sum_{v=1}^u |\omega_{j(v)}| \right)$, $S_{\alpha\lambda}(z)=\text{sign}(z)\max\left(|z|-\alpha\lambda, 0 \right)$
}
 Select $u^\ast_j$ as the smallest integer from $\left\{u: |\omega_{j(u)}|\geq \zeta_{ju} \geq |\omega_{j(u+1)}| \right\}$, where $\omega_{j(0)}=+\infty$,  $\omega_{j(\nu_j+1)}=0$\;
 Update ${\bm b}_j \leftarrow  \frac{ \zeta_{j{u^\ast_j}}\bm b_j}{C\|\bm b_j\|}$\;
\For{$k=1,\ldots,q_{1j}$}  
{\For{$l=1,\ldots,q_2$}  
 {Update ${W}_{1kl}^{(j)} \leftarrow \text{sign}\left({W}_{1kl}^{(j)} \right)\left(\zeta_{j{u^\ast_j}} \wedge \left|{W}_{1kl}^{(j)}\right| \right)$\;
 }}
\Output{$\bm b_j$ and $W_1^{(j)}$}
\caption{Updating $\bm b_j$ and $W_1^{(j)}$ in Model (\ref{eq:obj})}\label{Alg:W}
\end{algorithm}

\bibliographystyle{plainnat}
\bibliography{refs} 

\begin{thebibliography}{25}
\providecommand{\natexlab}[1]{#1}
\providecommand{\url}[1]{\texttt{#1}}
\expandafter\ifx\csname urlstyle\endcsname\relax
  \providecommand{\doi}[1]{doi: #1}\else
  \providecommand{\doi}{doi: \begingroup \urlstyle{rm}\Url}\fi

\bibitem[Berrendero et~al.(2016)Berrendero, Cuevas, and Torrecilla]{Berrendero:etal:16}
Jos{\'e}~R Berrendero, Antonio Cuevas, and Jos{\'e}~L Torrecilla.
\newblock Variable selection in functional data classification: a maxima-hunting proposal.
\newblock \emph{Statistica Sinica}, 7\penalty0 (2):\penalty0 619--638, 2016.

\bibitem[C\'{e}rou and Guyader(2006)]{Cerou:06}
F.~C\'{e}rou and A.~Guyader.
\newblock Nearest neighbor classification in infinite dimension.
\newblock \emph{ESAIM: Probability and Statistics}, 10:\penalty0 340--355, 2006.

\bibitem[Dai and Genton(2018)]{Dai:Genton:18}
Wenlin Dai and Marc~G Genton.
\newblock An outlyingness matrix for multivariate functional data classification.
\newblock \emph{Statistica Sinica}, 28\penalty0 (4):\penalty0 2435--2454, 2018.

\bibitem[Delaigle and Hall(2012)]{Delaigle:Hall:12}
Aurore Delaigle and Peter Hall.
\newblock Achieving near-perfect classification for functional data.
\newblock \emph{Journal of the Royal Statistical Society, Series~B}, 74:\penalty0 267--286, 2012.

\bibitem[He et~al.(2016)He, Zhang, Ren, and Sun]{He2016}
Kaiming He, Xiangyu Zhang, Shaoqing Ren, and Jian Sun.
\newblock Deep residual learning for image recognition.
\newblock \emph{In Proceedings of the IEEE Conference on Computer Vision and Pattern Recognition}, pages 770--778, 2016.

\bibitem[James and Hastie(2001)]{James:Hastie:01}
G.~M. James and T.~Hastie.
\newblock Functional linear discriminant analysis for irregularly sampled curves.
\newblock \emph{Journal of the Royal Statistical Society, Series~B}, 63:\penalty0 533--550, 2001.

\bibitem[LeCun et~al.(2015)LeCun, Bengio, and Hinton]{lecun2015deep}
Yann LeCun, Yoshua Bengio, and Geoffrey Hinton.
\newblock Deep learning.
\newblock \emph{nature}, 521\penalty0 (7553):\penalty0 436--444, 2015.

\bibitem[Lemhadri et~al.(2021)Lemhadri, Ruan, Abraham, and Tibshirani]{Lemhadri2018}
Ismael Lemhadri, Feng Ruan, Louis Abraham, and Robert Tibshirani.
\newblock Lassonet: A neural network with feature sparsity.
\newblock \emph{Journal of Machine Learning Research}, 130:\penalty0 10--18, 2021.

\bibitem[Li et~al.(2022)Li, Xiao, and Luo]{Li:etal:22}
Cai Li, Luo Xiao, and Sheng Luo.
\newblock Joint model for survival and multivariate sparse functional data with application to a study of alzheimer's disease.
\newblock \emph{Biometrics}, 78\penalty0 (2):\penalty0 435--447, 2022.

\bibitem[Lin and Jegelka(2018)]{Lin2018}
Hongzhou Lin and Stefanie Jegelka.
\newblock Resnet with one-neuron hidden layers is a universal approximator.
\newblock \emph{In Advances in Neural Information Processing Systems}, pages 6169--6178, 2018.

\bibitem[Moindji{\'e} et~al.(2024)Moindji{\'e}, Dabo-Niang, and Preda]{moindjie:etal:22}
Issam-Ali Moindji{\'e}, Sophie Dabo-Niang, and Cristian Preda.
\newblock Classification of multivariate functional data on different domains with partial least squares approaches.
\newblock \emph{Statistics and Computing}, 34\penalty0 (5), 2024.

\bibitem[M{\"u}ller(2005)]{Muller:05classification}
Hans-georg M{\"u}ller.
\newblock Functional modelling and classification of longitudinal data.
\newblock \emph{Scandinavian Journal of Statistics}, 32\penalty0 (2):\penalty0 223--240, 2005.

\bibitem[Rao and Reimherr(2023)]{rao2023nonlinear}
Aniruddha~Rajendra Rao and Matthew Reimherr.
\newblock Nonlinear functional modeling using neural networks.
\newblock \emph{Journal of Computational and Graphical Statistics}, 32\penalty0 (4):\penalty0 1248--1257, 2023.

\bibitem[Schmidt-Hieber(2020)]{Schmidt:19}
J.~Schmidt-Hieber.
\newblock Nonparametric regression using deep neural networks with relu activation function.
\newblock \emph{The Annals of Statistics}, 48\penalty0 (4):\penalty0 1875--1897, 2020.

\bibitem[Thind et~al.(2023)Thind, Multani, and Cao]{thind2023deep}
Barinder Thind, Kevin Multani, and Jiguo Cao.
\newblock Deep learning with functional inputs.
\newblock \emph{Journal of Computational and Graphical Statistics}, 32\penalty0 (1):\penalty0 171--180, 2023.

\bibitem[Wang et~al.(2013)Wang, Kim, and Li]{Wang:13}
Lan Wang, Yongdai Kim, and Runze Li.
\newblock Calibrating non-convex penalized regression in ultra-high dimension.
\newblock \emph{The Annals of Statistics}, 41\penalty0 (5):\penalty0 2505--2536, 2013.

\bibitem[Wang and Cao(2024)]{wang:cao:23}
Shuoyang Wang and Guanqun Cao.
\newblock Multiclass classification for multidimensional functional data through deep neural networks.
\newblock \emph{Electronic Journal of Statistics}, 18\penalty0 (1):\penalty0 1248--1292, 2024.

\bibitem[Wang et~al.(2023{\natexlab{a}})Wang, Cao, and Shang]{wang:etal:23}
Shuoyang Wang, Guanqun Cao, and Zuofeng Shang.
\newblock Deep neural network classifier for multi-dimensional functional data.
\newblock \emph{Scandinavian Journal of Statistics}, 50\penalty0 (4), 2023{\natexlab{a}}.

\bibitem[Wang et~al.(2023{\natexlab{b}})Wang, Huang, and Cao]{Wang:Huang:cao:23}
Shuoyang Wang, Yuan Huang, and Guanqun Cao.
\newblock Review on functional data classification.
\newblock \emph{WIREs Computational Statistics}, 16\penalty0 (1), 2023{\natexlab{b}}.

\bibitem[Wang et~al.(2023{\natexlab{c}})Wang, Shang, Cao, and Liu]{wang:etal:21a}
Shuoyang Wang, Zuofeng Shang, Guanqun Cao, and S.~Jun Liu.
\newblock Optimal classification for functional data.
\newblock \emph{Statistica Sinica}, 34\penalty0 (3), 2023{\natexlab{c}}.

\bibitem[Wang et~al.(2014)Wang, Nan, Zhu, and Koeppe]{Wang:etal:14:AOAS}
Xuejing Wang, Bin Nan, Ji~Zhu, and Robert Koeppe.
\newblock Regularized 3d functional regression for brain image data via haar wavelets.
\newblock \emph{The annals of applied statistics}, 8\penalty0 (2):\penalty0 1045, 2014.

\bibitem[Wang et~al.(2017)Wang, Nan, Zhu, Koeppe, and Frey]{Wang:etal:16:bios}
Xuejing Wang, Bin Nan, Ji~Zhu, Robert Koeppe, and Kirk Frey.
\newblock Classification of adni pet images via regularized 3d functional data analysis.
\newblock \emph{Biostatistics \& epidemiology}, 1\penalty0 (1):\penalty0 3--19, 2017.

\bibitem[Wu et~al.(2023)Wu, Beaulac, and Cao]{wu2023neural}
Sidi Wu, C{\'e}dric Beaulac, and Jiguo Cao.
\newblock Neural networks for scalar input and functional output.
\newblock \emph{Statistics and Computing}, 33\penalty0 (5):\penalty0 118, 2023.

\bibitem[Yao et~al.(2021)Yao, Mueller, and Wang]{yao2021deep}
Junwen Yao, Jonas Mueller, and Jane-Ling Wang.
\newblock Deep learning for functional data analysis with adaptive basis layers.
\newblock In \emph{International conference on machine learning}, pages 11898--11908. PMLR, 2021.

\bibitem[Yu et~al.(2023)Yu, Wade, Bondell, and Azizi]{Yu:etal:22}
Weichang Yu, Sara Wade, D.~Howard Bondell, and Lamiae Azizi.
\newblock Nonstationary gaussian process discriminant analysis with variable selection for high-dimensional functional data.
\newblock \emph{Journal of Computational and Graphical Statistics}, 32\penalty0 (2):\penalty0 588--600, 2023.

\end{thebibliography}

\end{document}